\documentclass[11pt]{article}

\usepackage{amsmath}
\usepackage{amssymb}
\usepackage{amsfonts}
\usepackage{cite}
\usepackage{bbm}
\usepackage[dvips]{graphicx}
\usepackage{subfigure}

\newtheorem{theorem}{Theorem}

\newtheorem{proposition}[theorem]{Proposition}
\newtheorem{lemma}[theorem]{Lemma}
\newenvironment{note}{\medskip\noindent\stepcounter{theorem}{\bf Remark~\arabic{theorem}}\,}
{\smallskip}
\newcommand{\Tr}{\operatorname{Tr}}
\newcommand{\vett}[1]{\boldsymbol{#1}}
\newcommand{\va}{\vett{\alpha}}
\newcommand{\vb}{\vett{\beta}}
\newcommand{\vg}{\vett{\gamma}}
\newcommand{\vd}{\vett{\delta}}
\newcommand{\tr}[1]{{\rm Tr}\big({#1}\big)}
\newcommand{\beq}{\begin{equation}}
\newcommand{\eeq}{\end{equation}}
\newcommand{\idmat}{\mathbbm{1}}
\newcommand{\idmap}{{\rm id}}
\newcommand{\bra}[1]{\langle #1 |}
\newcommand{\ket}[1]{| #1 \rangle}

\newcommand{\pro}[1]{\ket{#1}\bra{#1}}

\title{\bf A class of $2^N\times2^N$ bound entangled states revealed by non-decomposable maps}
\author{Marco Piani\\
\small Dipartimento di Fisica Teorica, Universit\`a di Trieste, Trieste, Italy\\
\small Istituto Nazionale di Fisica Nucleare, Sezione di Trieste, Trieste, Italy} 

\date{\null}

\begin{document}

\maketitle

\begin{abstract}
\noindent
We use some general results regarding positive maps to exhibit examples of non-decomposable maps and $2^N\times2^N$, $N\geq2$, bound entangled states, e.g. non distillable bipartite states of N+N qubits. 
\end{abstract}

\section{Introduction}
\label{sec:1}

Entanglement appears to be a basic resource in the fields of quantum information and quantum
computation (see \cite{NC,NC2} and references therein). Even if there is a sound definition of what an entangled state is \cite{Werner}, it is difficult to
determine if a given state is entangled or not.

There are different results in the literature regarding the 
classification of states. One of the more interesting \cite{Per,HHH1} is based on the use of linear maps which are positive (P) \cite{CH1,Sto,ZB} but not completely positive (CP) \cite{CH2,Tak,Kra}: we shall refer to them as PnCP maps. A map is P if it trasforms any state into another positive operator.  In the case of a bipartite system, a state is entangled if and only if there exists a PnCP map such that the operator obtained acting with the map on only one of the two subsystems is not positive any more. The simplest example of PnCP map is the operation of transposition $T$ (with respect to a given basis). The action of trasposition on one of the subsystems is called \emph{partial transposition} (PT). Because of the structure of the set of positive maps \cite{Sto,Wor}, in the $2\times2$ and $2\times3$ dimensional cases PT can ``detect'' all the entangled states: states that remain positive under PT (PPT states) are separable; states that develop negative eigenvalues under PT (NPT states) are entangled. Unfortunately in higher dimensions PT is not a ``complete'' test any more and there are PPT states which are entangled \cite{H1}.

The PnCP approach to the problem of entanglement characterization can also give information about the distillability of the state (see \cite{HHH0} for a review). A state is said to be distillable if, having at disposal a large number of copies of the state, it is possible to obtain some maximally entangled states, under the constraint of performing only local operations and  using classical communication. It turns out that a PPT entangled state (PPTES) can not be distilled, so that its entanglement can be considered ``bound'' \cite{HHH2};  however it can still be  useful for tasks that it would be impossible to perform classically \cite{HHH3}. In order to identify this bound entanglement it is necessary to use PnCP that are not decomposable, that is which can not be written as the sum of a CP map and a CP map composed with transposition. 

It is therefore clear that the study of P maps is strictly related to the study of entanglement, the link being provided by the Jamiolkowsky isomosphism \cite{Jam}. In this work we contribute to the phenomenology of positive maps \cite{CH3,Kye1,Kos} giving some general methods to construct classes of PnCP maps. In one simple instance we test their decomposability by finding at the same time examples of PPT (and therefore bound) entangled states of $N+N$ qubits.

In Section \ref{sec:2} we review some basic notions and results concerning the properties of 
positivity and complete positivity of maps and their relation to entanglement; we further give a method to construct particular classes of PnCP maps. 
In Section \ref{sec:3}, we use the results of Section \ref{sec:2} focusing on an example of PnCP map. To test the non-decomposability of this map we are lead quite naturally to define a set of states such that the condition of positivity under PT has a simple form. We then exhibit examples of $2^N\times2^N$ dimensional PPTES, thus proving that the map is non-decomposable.
 
\section{Linear maps and entanglement}
\label{sec:2}

We start with some basic facts about positive maps and entanglement, presented for finite $d$-dimensional systems $S_d$ described 
by the algebra of $d\times d$ matrices with complex entries $M_d({\mathbb C})$.
We shall denote 
by ${\cal S}_d$ the space of the states (density matrices), 
that is the convex set of positive $\rho\in M_d({\mathbb C})$ of unit trace.

The action of any hermiticity-preserving linear map 
$\Lambda:M_d({\mathbb C})\rightarrow M_d({\mathbb C})$ can be written as \cite{GKS}
\begin{equation}
\label{eq:linmap}
M_d({\mathbb C}) \ni X \mapsto\Lambda[X]=\sum_{k,i=0}^{d^2-1}\,
\lambda_{ki}\,F_k\, X \,F^\dagger_i\ ,
\end{equation}
where $F_k$'s are $d^2$ matrices $d\times d$,  forming an orthonormal basis in $M_d({\mathbb C})$ with respect to the Hilbert-Schmidt scalar product, $\tr{F^\dagger_i F_k}=\delta_{ki}$,
and $C_\Lambda=[\lambda_{ki}]$ is a generic hermitian matrix.
The map is also trace-preserving if and only if
$\sum_{k,i=0}^{d^2-1}\lambda_{ki}F_i^\dagger F_k=1$.

\begin{note}
Expression \eqref{eq:linmap} does not depend on the choice of the orthonormal basis
of the matrices $F_k$'s. In fact, let  $\{G_l\}$ be another orthonormal basis; there exists a $d^2\times d^2$ unitary matrix $U$, $U_{lk}=\Tr\big(G_l^\dagger F_k\big)$, such that $F_k=\sum_{l=0}^{d^2-1}\,U_{lk}\,G_l$.
Thus the action of $\Lambda$ can be written
\[
\Lambda[X]=\sum_{k,i=0}^{d^2-1}\,
\lambda_{lm}'\,G_l\, X \,G^\dagger_m\ 
\]
with $\lambda_{lm}'=(C_\Lambda')_{lm}=(UC_\Lambda U^\dagger)_{lm}$. On the other hand it is always possible to find an orthonormal basis $\{G_l\}$ such that $C_\Lambda$ is diagonal and \eqref{eq:linmap} reads 
$\Lambda[X]=\sum_{j=0}^{d^2-1}\,
\lambda_j\,G_j\, X \,G^\dagger_j$, where the $\lambda_j$'s are the eigenvalues of $C_\Lambda$. We will refer to such an orthonormal basis as a \emph{diagonal basis} for $\Lambda$.
\end{note}

Any linear map $\Lambda$ that is used to describe a physical state
transformation, must preserve the positivity of all states $\rho$,
otherwise the appearance of negative eigevalues in $\Lambda[\rho]$
would spoil its statistical interpretation which is based on the use
of the state eigenvalues as probabilities.

If a map preserves the positivity of the spectrum of all
$\rho$ we say it is \emph{positive}; however, it is not sufficient to make $\Lambda$
fully physically consistent.
Indeed, the system $S_d$ may always be thought to be statistically
coupled to an ancilla $n$-level system $S_n$. One is thus forced to
consider the action ${\rm id}_n\otimes\Lambda$ over the compound
system $S_n+S_d$, where by ${\rm id}_n$ we will denote in the following the identity action on
$M_n({\mathbb C})$. It is not only  $\Lambda$ that should be positive, but
also  ${\rm id}_n\otimes\Lambda$ for all $n$; 
such a property is called \emph{complete positivity} \cite{CH1,Tak,Kra}.
Complete positivity is necessary because
of the existence of entangled states of the compound system $S_n+S_d$,
namely of states that cannot be written as factorized linear convex 
combinations, that is as
\begin{equation}
\label{sepst}
\rho^{\rm sep}_{S_n+S_d}=\sum_{i}\,c_{i}\,\rho_{S_n}^{(i)}\otimes\rho_{S_d}^{(i)}\ ,\quad
c_{i}\geq0\ ,\quad\sum_{i}c_{i}=1\ .
\end{equation}
In fact, any PnCP map $\Lambda$, when acting partially as ${\rm id}\otimes\Lambda$, moves some entangled states 
out of the space of states; however, exactly for this reason, it may be used to detect entanglement \cite{Per,HHH1}.

In the first of the following two theorems we collect some results concerning positivity and complete
positivity; the second one is the Horodeckis' theorem on entanglement detection by positive maps.

In the space of states ${\cal S}_{d\times d}$ of the bipartite system $S_d+S_d$,
let us introduce the symmetric state
\begin{equation}
\label{pos2}
\vert\Psi^d_+\rangle=\frac{1}{\sqrt{d}}\sum_{j=1}^d\vert j\rangle\otimes\vert
j\rangle\ ,
\end{equation}
where $\vert j\rangle$, $j=1,2,\ldots,d$ is any fixed orthonormal
basis in ${\mathbb C}^d$,
and $P^d_+\equiv|\Psi^d_+\rangle\langle \Psi^d_+|$ $\in{\cal S}_{d\times d}$ is the corresponding projection onto it. 
\begin{theorem}
\label{TH1}
A linear map $\Lambda:M_d({\mathbb C})\rightarrow M_d({\mathbb C})$ is
\begin{itemize} 
\item[(i)]
positive if and only if
\begin{equation}
\label{pos}
\langle\phi\otimes\psi\vert\,
\big(\operatorname{id}_d\otimes\Lambda\big)[P^d_+]\,\vert\phi\otimes\psi\rangle=\frac{1}{d}
\langle\psi\vert\Lambda[\,\vert\phi^*\rangle\langle\phi^*\vert\,]
\vert\psi\rangle\geq 0
\end{equation}
for all (normalized) $\vert\phi\rangle\,,\,\vert\psi\rangle\,\in{\mathbb C}^d$, with
$\vert\phi^*\rangle$ denoting the conjugate of $\vert\phi\rangle$ with
respect to the fixed orthonormal basis in ${\mathbb C}^d$  \cite{Jam,K}.
\item[(ii)]
completely positive if and only if
\cite{CH1,HH}
\begin{equation}
\label{cpos}
\big(\operatorname{id}_d\otimes\Lambda\big)[P^d_+]\geq0\ .
\end{equation}
\item[(iii)]
completely positive
if and only if it can be expressed in the Kraus-Stinespring 
form~\cite{Tak,Kra} 
\begin{equation}
\label{cpos1}
\Lambda[\rho]=\sum_{j=0}^{d^2-1}\, K_j\,\rho\,K^\dagger_j\ ,
\end{equation} 
with $\sum_jK^\dagger_jK_j=1$ if $\Lambda$ is trace-preserving.
\end{itemize}
\end{theorem}
\begin{note}
It is evident that the properties of positivity and complete positivity depend on the coefficient matrix
$C_\Lambda$ of \eqref{eq:linmap}. In particular a linear map is CP if and only if $C_\Lambda$ is positive semidefinite. In fact in this case it is possible to obtain the Kraus-Stinespring of point (iii) of Theorem \ref{TH1} diagonalizing the coefficient matrix and using the fact that the eingenvalues of $C_\Lambda$ are positive:
\[
\Lambda[\rho]=\sum_{j=0}^{d^2-1}\, \lambda_j\,G_j\,\rho\,G^\dagger_j=
\sum_{j=0}^{d^2-1}\, \big(\sqrt{\lambda_j}\,G_j\big)\,\rho\,\big(\sqrt{\lambda_j}\,G_j\big)^\dagger,
\]
with $\sqrt{\lambda_j}\,G_j$ identified as the Kraus operator $K_j$.
\end{note}

\begin{theorem}
\label{pos6}
The following statements are equivalent \cite{HHH1}:
\begin{itemize}
\item[(i)] $\rho\in{\cal S}_{d\times d}$ is entangled;
\item[(ii)] for some positive map $\Lambda$ on $M_{d}({\mathbb C})$
\begin{equation}
D_\Lambda(\rho):=\Tr\Big(\big(\idmap_d\otimes\Lambda\big)[P^d_+]\,\rho\Bigr)\,<\,0;
\end{equation}
\item[(iii)] for some positive map $\Lambda$ on $M_{d}({\mathbb C})$
\[
\rho'=(\idmap_d\otimes\Lambda\big)[\rho]
\]
is not a positive operator any more.
\end{itemize}
\end{theorem}

\begin{note}
For any $\ket{\psi}=\sum_{i,j=1}^d\psi_{ij}\ket{i}\otimes\ket{ j}$ in ${\mathbb C}^{d\times d}$
there is $A_\psi$ in $M_{d}({\mathbb C})$ such that $\bra{j}A\ket{i}=\psi_{ij}$ and therefore
$\ket{\psi}=\sqrt{d}(\idmat_d\otimes A_\psi)\ket{\Psi^d_+}$. It is then clear that for any state $\rho\in{\cal S}_{d\times d}$ there is a CP map $\Lambda_\rho$, characterized by a coefficient matrix $C_{\Lambda_\rho}$, such that $\rho=\big(\idmap_d\otimes\Lambda_\rho\big)[P^d_+]$. Thus for any P map $\Lambda$ characterized by a coefficient matrix $C_{\Lambda}$, we have
\[
D_\Lambda(\rho)=\frac{1}{d^2}\Tr\big(C_\Lambda C_{\Lambda_\rho}\big)
\]
with the two coefficient matrices expressed in the same orthonormal basis.
\end{note}

\begin{note}
\label{rem:psiplus}
We note two interesting properties of the symmetric state $\ket{\Psi^d_+}$:
\begin{enumerate}
\item for all matrices $A$, $B$ acting on ${\mathbb C}^d$ one has
\[
A\otimes B\ket{\Psi^d_+}=\idmat_d\otimes BA^T\ket{\Psi^d_+}=AB^T\otimes\idmat_d\ket{\Psi^d_+};
\]
\item under partial transposition $P^d_+$ gives raise to the \emph{flip operator}
\[
V:=\sum_{i,j=1}^d\ket{i}\bra{j}\otimes\ket{j}\bra{i}=d(\idmap_d\otimes T)[P^d_+]
\]
which is such that
\[
V\vert\psi\otimes\phi\rangle=\vert\phi\otimes\psi\rangle\ ,\quad
V\Bigl(A\otimes B\Bigr)V=B\otimes A \ .
\]
\end{enumerate}
\end{note}

Unlike the case of CP maps, there is no general prescription
on $C_\Lambda$ 
ensuring that $\Lambda$ preserves the positivity of $\rho$.
For instance, if $C_\Lambda$ is not positive, then, by separating positive 
and negative eigenvalues, one sees that every
$\Lambda$ can be written as the difference of two CP maps
$\Lambda_{1,2}$ \cite{Yu}:
\begin{equation}
\label{pos3}
\Lambda[\rho]=\sum_{\lambda_j\geq0}\lambda_j\,G_j\,\rho\,G_j^\dagger\ -\
\sum_{\lambda_j<0}|\lambda_j|\,G_j\,\rho\,G_j^\dagger.
\end{equation} 
with $G_j$ a diagonal basis.
However, no general rule is known that may allow us to recognize the
positivity of $\Lambda$ by looking at the eigenvalues $\lambda_j$ and at
the matrices $G_j$.  From the point (i) of Theorem \ref{TH1} and from \eqref{pos3}, it easy to derive that
one should check the positivity of
\begin{equation}
\langle\psi|\Lambda[|\phi\rangle\langle\phi|]|\psi\rangle=
 \sum_{\lambda_j\geq0}\lambda_j\,|\langle\psi|G_j|\phi\rangle|^2 -\
\sum_{\lambda_j<0}\lambda_j\,|\langle\psi|G_j|\phi\rangle|^2
\end{equation}
for all (normalized) $\vert\phi\rangle\,,\,\vert\psi\rangle\,\in{\mathbb C}^d$.

In the following Theorem we will give a sufficient condition for positivity of a class of maps
on $M_{d_1}({\mathbb C})\otimes M_{d_2}({\mathbb C})$.

\begin{theorem}
\label{TH2}
Let $\Lambda_i$ be maps acting on $M_{d_i}({\mathbb C})$, $i=1,2$, in the following way
\beq
\label{eq:lambdaherm}
\Lambda_i[X]=\sum_{\mu=0}^{d_i^2-1}\,\lambda^{(i)}_\mu\,F^{(i)}_\mu\,X\,F^{(i)}_\mu,
\eeq
i.e. they admit hermitian diagonal bases
\[
F^{(i)}_\mu=\big(F^{(i)}_\mu\big)^\dagger
\]
for all $\mu=0,\ldots,d_i^2-1$, $i=1,2$.
If all the coefficients $\lambda^{(i)}_\mu$ are positive apart from one, let us say
$\lambda^{(2)}_k=-|\lambda^{(2)}_k|$, and all the positive coefficients are greater or equal to $|\lambda^{(2)}_k|$, then the map $\Lambda:M_{d_1\times d_2}({\mathbb C})\rightarrow M_{d_1\times d_2}({\mathbb C})$,
\beq
\label{eq:sumlambda}
\Lambda=\Lambda_1\otimes\idmap_{d_2}+\idmap_{d_1}\otimes\Lambda_2
\eeq
is positive.
\end{theorem}
\smallskip
\noindent
{\bf Proof} \quad We have to check that
\[
D(\phi,\psi):=\langle\psi|\Lambda[|\phi\rangle\langle\phi|]|\psi\rangle\geq0
\]
for all $\vert\phi\rangle\,,\,\vert\psi\rangle\,\in{\mathbb C}^{d_1\times d_2}$, which we can both be expanded on a basis \mbox{$\big\{\ket{i}\otimes\ket{j}\big\}$}:
\begin{equation}
\vert\phi\rangle=\sum_{i=1}^{d_1}\sum_{j=1}^{d_2}\Phi_{ij}\ket{i}\otimes\ket{j},\qquad
\ket{\psi}=\sum_{k=1}^{d_1}\sum_{l=1}^{d_2}\Psi_{kl}\ket{k}\otimes\ket{l},
\end{equation}
so that they are determined by the coefficient matrices $\Phi,\,\Psi$.
It is straighforward to find the following expression for $D(\phi,\psi)$:
\beq
\label{eq:Dphipsi}
D(\phi,\psi)=\sum \lambda_\mu^{(1)} \Big|\Tr\Big(F^{(1)}_\mu\Phi\Psi^\dagger\Big)\Big|^2
+
\sum \lambda_\nu^{(2)} \Big|\Tr\Big(F^{(2)}_\nu\big(\Psi^\dagger\Phi\big)^T\Big)\Big|^2
\eeq
Since $\{F^{(1)}_\mu\}_{\mu=0}^{d_1^2-1}$ and $\{F^{(2)}_\nu\}_{\mu=0}^{d_2^2-1}$ are two orthonormal bases and $\Tr(A^T)=\Tr(A)$, we have
\[
\sum_\mu\Big(\Tr\Big(F^{(1)}_\mu\Phi\Psi^\dagger\Big)\Big)^2=\Tr\big(\Phi\Psi^\dagger\Phi\Psi^\dagger\big)=\sum_\nu\Big(\Tr\Big(F^{(2)}_\nu\big(\Psi^\dagger\Phi\big)^T\Big)\Big)^2
\] 
and, using the triangle inequality, we have
\beq
\label{eq:triangleineq}
\Big|\Tr\Big(F^{(2)}_k\big(\Psi^\dagger\Phi\big)^T\Big)\Big|^2\leq
\sum_\mu\Big|\Tr\Big(F^{(1)}_\mu\Phi\Psi^\dagger\Big)\Big|^2+
\sum_{\nu\neq k}\Big|\Tr\Big(F^{(2)}_\nu\big(\Psi^\dagger\Phi\big)^T\Big)\Big|^2
\eeq
From the hypothesis of the theorem, the above inequality and \eqref{eq:Dphipsi} we find
\begin{equation}
\begin{split}
D(\phi,\psi)
\geq
|\lambda^{(2)}_k|
\Bigg[
&\sum_\mu\Big|\Tr\Big(F^{(1)}_\mu\Phi\Psi^\dagger\Big)\Big|^2\\
+&\sum_{\nu\neq k}\Big|\Tr\Big(F^{(2)}_\nu\big(\Psi^\dagger\Phi\big)^T\Big)\Big|^2
-
\Big|\Tr\Big(F^{(2)}_k\big(\Psi^\dagger\Phi\big)^T\Big)\Big|^2
\Bigg]
\geq 0.
\end{split}
\end{equation}
for all $\vert\phi\rangle\,,\,\vert\psi\rangle$; therefore $\Lambda$ is P.
\hfill$\blacksquare$
\medskip

\begin{note}
We have just shown that any map $\Lambda$ of the form \eqref{eq:sumlambda} is positive. It is moreover PnCP as soon as its matrix of coefficients is not positive, i.e. as soon as the negative contribution in $\idmap_{d_1}\otimes\Lambda_2$ due to $\lambda^{(2)}_k=-|\lambda^{(2)}_k|$ is not actually cancelled by 
terms in $\Lambda_1\otimes\idmap_{d_2}$.
\end{note}

\begin{note}
The previous theorem is suggested by a similar result regarding dynamical semigroups \cite{BFP2,BFPosid}. A \emph{dynamical semigroup} is a set of hermiticity and trace preserving linear maps $\gamma_t$, $t\geq 0$, on ${\cal S}_d$ which obey a semigroup composition
law $\gamma_t\circ\gamma_s=\gamma_{t+s}$, for any $t,s\geq0$. Semigroups are used to describe the dynamics of a 
system immersed in an environment and weakly coupled to it \cite{Spo,AL,BP,BF0}. With the further assumption of continuity in $t$ (time) the semigroup has the form $\gamma_t=\exp(tL)$, where $L$ is a map called the \emph{generator} which determines all the properties of the semigroup. The issue of complete positivity in the description of the evolution of dynamical systems is indeed related to the existence of entangled states \cite{BFBa,BFR3,BFR4,BFP1,BFPR}.
The generator of a factorized semigroup $\exp(tL)=\exp(tL_1)\otimes\exp(tL_2)$ on ${\cal S}_{d_1}\otimes{\cal S}_{d_2}$  is $L=L_1\otimes\idmap_{d_2}+\idmap_{d_1}\otimes L_2$, which is similar to \eqref{eq:sumlambda}.
\end{note}

Given a set of positive maps $\mathcal{P}=\{\Lambda_1, \ldots,\Lambda_p\}$ we can define a larger set of positive maps
\beq
\Omega(\mathcal{P})=\Big\{\sum_{i=0}^p\Gamma_i\circ\Lambda_i\,\Big|\,\Gamma_i \textrm{ CP}\Big\},
\eeq
with $\Lambda_0=\idmap$.
Then, given a set of PnCP maps $\big\{\Lambda_i^{\rm PnCP}\big\}_i$, we can conctruct a whole class $\Omega\big(\big\{\Lambda_i^{\rm {PnCP}}\big\}\big)$ of P maps, potentially PnCP. It is quite evident that no map in $\Omega\big(\big\{\Lambda_i^{\rm {PnCP}}\big\}_i\big)$ gives a stronger test for entanglement, in the sense of Theorem \ref{pos6}, than the ensamble of tests performed with the single $\Lambda_i^{\rm {PnCP}}$'s. In particular, if a map is in $\Omega\big(\{T\}\big)$, it is said to be \emph{decomposable} and cannot provide a stronger test than PT.
According to a theorem by Woronowicz~\cite{Wor},
all P maps $M_2({\mathbb C})\rightarrow M_2({\mathbb C})$ are decomposable, whence
the transposition detects all the entangled states in
${\cal S}_{2\times 2}$; in other words,  $\big(\idmap_2\otimes T_2\big)[\rho]$ is
non-positive if and only if $\rho$ is entangled.
On the contrary, when $d\geq 3$, there are PPT states which are entangled (PPTES)
\cite{HHH0,CH2,Sto,H1}.
The entanglement in a PPTES can not be distilled by
means of local operations and classical communication \cite{HHH2}, 
therefore it is referred to as bound-entanglement.

The relation between non-decomposability of maps and PPT entangled states is summarized in the following proposition:

\begin{proposition}
\label{th:proposindec}
If $\Lambda$ is positive on ${\cal S}_d$,  
$\rho\in{\cal S}_{d\times d}$ is PPT and $D_\Lambda(\rho)<0$, then
$\Lambda$ is not decomposable and $\rho$ is PPTES.
\end{proposition}

\section{A class of $2^N\times2^N$ bound entangled states}
\label{sec:3}

We want to use the results of Theorem \ref{TH2}. We notice that $\big\{\sigma_\mu/\sqrt{2}\big\}_{\mu=0}^3$, with $\sigma_0$ the 2-dimensional identity matrix and $\sigma_i$, $i=1,2,3$, the Pauli matrices, is a hermitian orthonormal basis in $M_{2}({\mathbb C})$.
Let $L^{(k)}$ be the the set
\[
L^{(k)}:=\Bigl\{(\omega_1,\ldots,\omega_k)\,\Big|\,\omega_i=0,1,2,3\,,\,i=1,\ldots,k\,\Bigr\}
\]
whose elements are $k$-dimensional (integer) vectors $\vett{\omega}$. Let us take $m\geq n$ and let $L$ be the lattice 
\begin{equation}
L:=L^{(m)}\times L^{(n)}=\Bigl\{(\va,\vb)\,\Big|\,\va\in L^{(m)},\vb\in L^{(n)}\Bigr\}
\end{equation}
 with $4^N,\,N=m+n$ elements. 
 In a geometric representation $L$ can be considered in an $N$-dimensional integer space as  a hypercube whose sides contain 4 points. Every index among $\alpha_1,\dots,\alpha_m,\beta_1,\ldots,\beta_n$ is then a coordinate. 
 We will associate to the points of $L$ the tensor products of Pauli matrices
\begin{equation}
\sigma_{\vett{\alpha}\vett{\beta}}:=\sigma_{\vett{\alpha}}\otimes\sigma_{\vett{\beta}}
						=(\,\sigma_{\alpha_1}\otimes\cdots\otimes\sigma_{\alpha_m}\,)\,
							\otimes \, (\,\sigma_{\beta_1}\otimes\cdots\otimes\sigma_{\beta_n}\,).
\end{equation}
 It is clear that
$\bigg\{\sigma_{\vett{\alpha}}/{(\sqrt{2})}^m\bigg|\va\in L^{(m)}\bigg\}$ and $\bigg\{\sigma_{\vett{\beta}}/{(\sqrt{2})}^n\bigg|\vb\in L^{(n)}\bigg\}$
are orthonormal hermitian bases respectively in $M_{2^m}({\mathbb C})$ and $M_{2^n}({\mathbb C})$,
while
\[
\Sigma=\bigg\{\sigma_{\vett{\alpha}\vett{\beta}}/(\sqrt{2})^N\,\bigg|\,(\vett{\alpha},\vett{\beta})\in L\bigg\}
\]
is an orthonormal hermitian basis in $M_{2^m}({\mathbb C})\otimes M_{2^n}({\mathbb C})\simeq M_{2^{N}}({\mathbb C})$.

Let us consider the map
\beq
\label{eq:indecmap}
\Lambda_{\vb_0}=\Lambda_1\otimes\idmap_{2^n}+\idmap_{2^m}\otimes\Lambda_2,
\eeq
with
\begin{align*}
\Lambda_1[X]&=\frac{1}{2^n}\sum_{\va\in L^{(m)}}\,\frac{\sigma_{\vett{\alpha}}}{\sqrt{2^m}}\,X\,\frac{\sigma_{\vett{\alpha}}}{\sqrt{2^m}}\\
\Lambda_2[X]&=\frac{1}{2^m}\Bigg(\sum_{\vb\,\in \,L^{(n)}\backslash\{\vb_0\}}\,\frac{\sigma_{\vb}}{\sqrt{2^n}}\,X\,\frac{\sigma_{\vb}}{\sqrt{2^n}}\,-\,\frac{\sigma_{\vb_0}}{\sqrt{2^n}}\,X\,\frac{\sigma_{\vb_0}}{\sqrt{2^n}}\Bigg).
\end{align*}
and $\vb_0\neq\vett{0}_m$, denoting with $\vett{0}_k$ the $k$-dimensional null vector $(0,\ldots,0)\in L^{(k)}$.
Since $m\geq n$, the map $\Lambda_{\vb_0}$ satisfies the hypothesis of Theorem \ref{TH2}  and is therefore P. Note that $\idmap_{2^k}[X]=\sigma_{\vett{0}_k}\,X\,\sigma_{\vett{0}_k}$. In the basis $\Sigma$ the coefficient matrix $C_{\Lambda_{\vb_0}}$ is diagonal with eigenvalues
\beq
\label{eq:evLambda}
\lambda_{\va\vb}=\begin{cases}
2 & (\va,\vb)=(\vett{0}_m,\vett{0}_n)\\
1 & \Big((\va=\vett{0}_m)\land(\vb\neq\vett{0}_n,\vb_0)\Big)\,\lor\, \Big((\va\neq\vett{0}_m)\land(\vb=\vett{0}_n)\Big)\\
-1 & (\va,\vb)=(\vett{0}_m,\vb_0)\\
0 & (\va\neq\vett{0}_m)\land(\vb\neq\vett{0}_n).
\end{cases}
\eeq
It is therefore clear that, because of our choice for $\vb_0$, the map $\Lambda_{\vb_0}$ is PnCP.  Accordingly to Proposition \ref{th:proposindec}, we will show that it is also non-decomposable   exhibiting a PPT state $\rho$ such that $D_{\Lambda_{\vb_0}}(\rho)<0$.

We construct orthogonal one-dimensional projectors
\beq
\label{sqlatst1}
P_{\va\vb}=\pro{\psi_{\va\vb}}\qquad \ket{\psi_{\va\vb}}:=\idmat_{2^N}\otimes\sigma_{\vett{\alpha}\vett{\beta}}\,\ket{\Psi_+^{2^N}}
\eeq
such that
\[
P_{\vett{\alpha}\vett{\beta}}\,P_{\vett{\gamma}\vett{\varepsilon}}=\,
\delta_{\vett{\alpha}\vett{\gamma}}\,\delta_{\vett{\beta}\vett{\varepsilon}}\ P_{\vett{\alpha}\vett{\beta}}
\]
The states $\ket{\psi_{\va\vb}}$, $(\va,\vb)\in L$, are $4^N$ maximally entagled states forming an orthonormal basis in
${\mathbb C}^{2^N}\otimes{\mathbb C}^{2^N}$.

We shall call \emph{lattice states} (LS) the states diagonal in the $\{\ket{\psi_{\va\vb}}\}$ basis, that is the mixtures $\rho_\pi$ belonging to
the convex span of the projectors
$P_{\vett{\alpha}\vett{\beta}}$:
\begin{equation}
\label{sqlatst3}
\rho_\pi:=\sum_{(\vett{\alpha},\vett{\beta})\in L} \pi_{\vett{\alpha}\vett{\beta}}\, \,P_{\vett{\alpha}\vett{\beta}}
\ ,\quad \pi_{\vett{\alpha}\vett{\beta}}\geq0\ ,\ \sum_{(\vett{\alpha},\vett{\beta})\in
L}\pi_{\vett{\alpha}\vett{\beta}}=1\ .
\end{equation}

We now analyze the problem of deciding which $\rho_\pi$ are PPT.
We start by operating the partial transposition on $P_{\vett{\alpha}\vett{\beta}}$ in~(\ref{sqlatst1}), obtaining 
\begin{equation}
\label{ptr1}
\rho_\pi^{T_2}:={\rm id}_{2^N}\otimes T[\rho_I]=
\frac{1}{2^N}\,\sum_{(\vett{\alpha},\vett{\beta})\in L}\,\pi_{\vett{\alpha}\vett{\beta}}\,V_{\vett{\alpha}\vett{\beta}},
\eeq
with
\beq
V_{\vett{\alpha}\vett{\beta}}:=\idmat_{2^N}\otimes\sigma_{\vett{\alpha}\vett{\beta}}
\,V\,\idmat_{2^N}\otimes\sigma_{\vett{\alpha}\vett{\beta}}\ .
\end{equation}

\begin{lemma} 
The spectral decompostion of $V_{\vett{\alpha}\vett{\beta}}$ is
\[
V_{\vett{\alpha}\vett{\beta}}=\sum_{(\vett{\gamma},\vett{\delta})\in
L}\Xi_{\vett{\alpha}\vett{\gamma}}\,\Xi_{\vett{\delta}\vett{\beta}}\, P_{\vett{\gamma}\vett{\delta}}
\]
with
\[
\Xi_{\vett{\alpha}\vett{\gamma}}=
\prod_{i=1}^m\Xi_{\alpha_i\gamma_i}
\qquad
\Xi_{\alpha\gamma}=(-1)^{\delta_{|\alpha-\gamma|,2}}.
\]
\end{lemma}

\noindent
{\bf Proof}\quad
From Remark \ref{rem:psiplus} and $V\vert\Psi^{2^N}_+\rangle=\vert\Psi_+^{2^N}\rangle$, 
it follows that
\begin{eqnarray*}
V_{\vett{\alpha}\vett{\beta}}\vert\Psi_{\vett{\gamma}\vett{\delta}}\rangle&=&
\prod_{i=1}^m\varepsilon_{\alpha_i}\varepsilon_{\gamma_i}
\prod_{j=1}^n\varepsilon_{\beta_j}\varepsilon_{\delta_j}\idmat_{2^N}\otimes\Biggl(\Bigl(\sigma_{\vett{\alpha}}\sigma_{\vett{\gamma}}\sigma_{\vett{\alpha}}\Bigr)
\otimes\Bigl(\sigma_{\vett{\beta}}\sigma_{\vett{\delta}}\sigma_{\vett{\beta}}\Bigr)\Biggr)
\vert\Psi_+^{2^N}\rangle\\
&=&
\prod_{i=1}^m\Bigl(\varepsilon_{\alpha_i}\varepsilon_{\gamma_i}\eta_{\alpha_i\gamma_i}\Bigr)\
\prod_{j=1}^n\Bigl(\varepsilon_{\beta_j}\varepsilon_{\delta_j}\eta_{\beta_j\delta_j}\Bigr)
\ \vert\Psi_{\vett{\gamma}\vett{\delta}}\rangle\ ,
\end{eqnarray*}
where
\[
\eta_{\alpha\gamma}:=\begin{cases}1& \alpha=0\vee\gamma=0\vee\alpha=\gamma\\
							-1&\alpha\neq0\wedge\gamma\neq0\wedge\alpha\neq\gamma
				\end{cases}\\
\]
according to the table
$$
\hbox{\begin{tabular}{ l || r | r | r | r }
$\gamma\backslash\alpha$&0&1&2&3\\ 
\hline\hline
0&1&1&1&1\\ \hline
1&1&1&-1&-1\\ \hline
2&1&-1&1&-1\\ \hline
3&1&-1&-1&1
\end{tabular}}\ ,
$$
and $\varepsilon_\alpha=(-1)^{\delta_{\alpha,2}}$.
Setting 
$\Xi_{\gamma\alpha}=\varepsilon_\alpha\varepsilon_\gamma\,\eta_{\alpha\gamma}$,
the result follows by direct inspection.
\hfill$\blacksquare$

\begin{theorem}
\label{TH4}
A lattice state remains positive under partial transposition if and only if, for any lattice site $(\vett{\alpha},\vett{\beta})\in L$,
\begin{equation}
\label{eq:Ipi}
J_{\vett{\alpha}\vett{\beta}}=\sum_{a=0}^{m+n}\,(-1)^a\,\Bigg(\sum_{(\vett{\gamma},\vett{\delta})\in L_{\vett{\alpha}\vett{\beta}}^a}\pi_{\vett{\gamma}\vett{\delta}}\Bigg)\geq\,0,
\end{equation}
where $L_{\vett{\alpha}\vett{\beta}}^a\subseteq L$ is the set of lattice points $(\vett{\gamma},\vett{\delta})$ such that exactly $a$ indices among $\gamma_1,\ldots,\gamma_m,\delta_1,\ldots,\delta_n$ are equal to the corresponding indices among
$\alpha_1,\ldots,\alpha_m,\beta_1,\ldots,\beta_n$.
\end{theorem}

\noindent
{\bf Proof}\quad
Using the previous Lemma, from \eqref{ptr1} we have
\begin{equation}
\rho_\pi^{T_1}=\frac{1}{2^N}\sum_{(\vett{\alpha},\vett{\beta})\in L}\Bigg(
\sum_{(\vett{\gamma},\vett{\delta})\in L}\pi_{\vett{\gamma}\vett{\delta}}\Xi_{\vett{\alpha}\vett{\gamma}}
\Xi_{\vett{\beta}\vett{\delta}}
\Bigg)
P_{\vett{\alpha}\vett{\beta}}.
\end{equation}
and therefore $\rho_\pi^{T_1}$ is positive if and only if
\begin{equation}
\label{condPPT}
\sum_{(\vett{\gamma},\vett{\delta})\in L}\pi_{\vett{\gamma}\vett{\delta}}\Xi_{\vett{\alpha}\vett{\gamma}}
\Xi_{\vett{\beta}\vett{\delta}}\geq0\quad\forall(\vett{\alpha},\vett{\beta})\in L.
\end{equation}
We introduce the bijection
$L\rightarrow L$ given by 
$(\vett{\alpha},\vett{\beta})\mapsto(\vett{\widetilde{\alpha}},\vett{\widetilde{\beta}})$, where
$\vett{\widetilde{\alpha}}=(\widetilde{\alpha}_1,\ldots,\widetilde{\alpha}_m)$ and $\widetilde{\mu}:=(\mu+2)\!\!\mod(4)$, $0\leq\mu\leq 3$. It then follows that $\Xi_{\alpha\gamma}=(-1)^{\delta_{\alpha,\widetilde{\gamma}}}$.  The PPT condition can therefore be written as
\[
\sum_{\vett{\gamma}\vett{\delta}\in L}\pi_{\vett{\gamma}\vett{\delta}}
	\prod_{i=1}^m(-1)^{\delta_{\widetilde{\alpha}_i\gamma_i}}
	\prod_{j=1}^n(-1)^{\delta_{\widetilde{\beta}_j\delta_j}}\geq0,
\]
whence, since it must hold for all $(\vett{\alpha},\vett{\beta})$, there must be
\begin{equation}
\label{eq:Iab}
J_{\vett{\alpha}\vett{\beta}}:=\sum_{\vett{\gamma}\vett{\delta}\in L}\pi_{\vett{\gamma}\vett{\delta}}
	\prod_{i=1}^m(-1)^{\delta_{\alpha_i\gamma_i}}
	\prod_{j=1}^n(-1)^{\delta_{\beta_j\delta_j}}\geq0.
\end{equation}
We now split the sum over $\vg,\,\vd$ into different sums according to the number of $\delta$ conditions that are satisfied, that is we isolate the contributions of different $L_{\vett{\alpha}\vett{\beta}}^a$; explicitly
\[
J_{\vett{\alpha}\vett{\beta}}=\sum_{a=0}^{m+n}\sum_{(\vett{\gamma},\vett{\delta})\in L_{\vett{\alpha}\vett{\beta}}^a}\pi_{\vett{\gamma}\vett{\delta}}\,
\prod_{i=1}^m(-1)^{\delta_{\alpha_i\gamma_i}}
	\prod_{j=1}^n(-1)^{\delta_{\beta_j\delta_j}}\geq0.
\]
The theorem follows noticing that
\[
(\vett{\gamma},\vett{\delta})\in L_{\vett{\alpha}\vett{\beta}}^a
\quad
\Rightarrow
 \quad
 \prod_{i=1}^m(-1)^{\delta_{\alpha_i\gamma_i}}\prod_{j=1}^n(-1)^{\delta_{\beta_j\delta_j}}=(-1)^a
\]
\hfill$\blacksquare$
\medskip

The previous condition for positive partial transposition is necessary and sufficient on the class of lattice states. For the sake of simplicity we now focus on a subset of these states. We will call \emph{equidistributed LS} (ELS) the LS such that
\[
\pi_{\va\vb}=
\begin{cases}
\frac{1}{N_I}	& (\vett{\alpha},\vett{\beta})\in I \\
0			& (\vett{\alpha},\vett{\beta})\notin I
\end{cases}
\]
with $I\subseteq L$ a subset of $L$ and $N_I:=\hbox{card}(I)$, that is states
\begin{equation}
\label{eqsqlatst}
\rho_I=\frac{1}{N_I}\sum_{(\vett{\alpha},\vett{\beta})\in I}\, P_{\vett{\alpha}\vett{\beta}}.
\end{equation}
Such states are completely characterized by a set $I\subseteq L$. The condition of positivity under PT \eqref{eq:Ipi} becomes
\[
J_{\vett{\alpha}\vett{\beta}}=\frac{1}{N_I}\sum_{a=0}^{m+n}\,(-1)^a\, N_{L_{\vett{\alpha}\vett{\beta}}^a}\geq\,0
\]
where $N_{L_{\vett{\alpha}\vett{\beta}}^a}=\hbox{card}\big({L_{\vett{\alpha}\vett{\beta}}^a}\big)$

It is in principle possible to construct all the ELS that are PPT. A similar task as been accomplished in \cite{BFPosid} for the case $m=n=1,\,N=2$, i.e. for the ELS  $\rho_I\in{\cal S}_{4\times4}$. In the present work, instead, we just show that for any $N\geq2$ among the ELS there is \emph{at least} a PPTES.

We first need the following lemma.

\begin{lemma}
\label{TH5}
The ELS described by
\beq
\label{eq:pointdistribution}
I_C=\big\{(\va,\vb)\,\big|\,\alpha_i\neq0\,\land\,\beta_j\neq0\quad\textrm{for }i=1,\ldots, m,\,j=1,\ldots,n\big\},
\eeq
$N_{I_C}=3^N$, is positive under partial transposition and $J_{\vett{\alpha}\vett{\beta}}\geq\frac{1}{N_{I_C}}$ for all $(\va,\vb)\in L$.
\end{lemma}

\noindent
{\bf Proof}\quad
Consider first the case $(\va,\vb)\in I_C$. Then
\beq
\label{eq:cubegII}
J_{\vett{\alpha}\vett{\beta}}=\frac{1}{3^N}\sum_{a=0}^N(-1)^a\binom{N}{a} g_a
\eeq
where the coefficient $g_a$ is such that $\binom{N}{a}g_a$ is the number of points in $L_{\vett{\alpha}\vett{\beta}}^a$, that is of points satisfying exactly $a$ conditions, as expressed by the $\delta$'s appearing, for example, in \eqref{eq:Iab}. To show the validity of \eqref{eq:cubegII}, let us denote by:
\begin{itemize}
\item $A(a)$ a set of $a$ conditions of the form ``$\alpha_i=\gamma_i$'', that is, equivalently, of $a$ numbers chosen between $\{1,2,\ldots,N\}$; 
\item $G_{A(a)}\subseteq I_C$ the set of points satisfying conditions $A(a)$ and no further ones.
\end{itemize}
We now notice that:
\begin{itemize}
\item $\binom{N}{a}$ is the number of different ways to choose $a$ conditions among $N$, that is the number of different sets $A(a)$;
\item given two sets $A(a),\,A'(a)$, if $A(a)\neq A'(a)$ then $G_{A(a)}$ and $G_{A'(a)}$ are disjoint, $G_A\cap G_{A'}=\emptyset$;
\item two sets of points $G_{A(a)}$, $G_{A'(a)}$ satisfying different conditions $A(a)$, $A'(a)$ are mapped one into the other by suitable permutations of the indices. Thus, they must contain the same number of points: $\hbox{card}(G_A)=g_a$ for all $A(a)$.
\end{itemize}
In summary, for each $a$, the set  $L_{\vett{\alpha}\vett{\beta}}^a$ can be split into $\binom{N}{a}$ disjointed sets $G_A(a)$ each containing $g_a$ points.\\
We claim that  $g_a=2^{N-a},\,a=0,\ldots,N$. This is certainly true for $a=N$, since there is only one point satisfying $N$ conditions. Let us suppose the statement be true for $a=\tilde{a},\ldots,N$. Then $g_{\tilde{a}-1}$ is given by
\beq
\label{eq:g_a_II}
\begin{split}
g_{\tilde{a}-1}&=3^{N-(\tilde{a}-1)}-\sum_{i=1}^{N-(\tilde{a}-1)}\binom{N-(\tilde{a}-1)}{i}g_{(\tilde{a}-1)+i}\\
		&=3^{N-(\tilde{a}-1)}-\sum_{i=1}^{N-(\tilde{a}-1)}\binom{N-(\tilde{a}-1)}{i}2^{N-[(\tilde{a}-1)+i]}\\
		&=3^{N-(\tilde{a}-1)}-\sum_{i=0}^{N-(\tilde{a}-1)}\binom{N-(\tilde{a}-1)}{i}2^{[N-(\tilde{a}-1)]-i]}
				+2^{N-(\tilde{a}-1)}\\
		&=2^{N-(\tilde{a}-1)}.
\end{split}
\eeq 
The relation between the $g_a$'s written in the first line of \eqref{eq:g_a_II} is easily explained: a set of points satisfying $\tilde{a}-1$  conditions (and no further) is given by the number of points satisfying \emph{at least} $\tilde{a}-1$ conditions minus all the disjoint sets satisfying exactly $i$ further conditions chosen among the remaining $N-(\tilde{a}-1)$.\\
Therefore
\beq
J_{\va\vb}=\frac{1}{3^N}\sum_{a=0}^N\binom{N}{a}(-1)^a 2^{N-a}=\frac{1}{3^N}
\eeq

For the state to be PPT, the condition $J_{\va\vb}\geq0$ must hold for any choice of $(\va,\vb)\in L$. We have already considered the case $(\va,\vb)\in I_C$. We now show this to be the worst case, in the sense that $J_{\va\vb}$ is the smallest possible. In fact consider the case where $k$ indices $\mu_{i_1},\ldots,\mu_{i_k}$ among the indices $(\vett{\gamma},\vett{\delta})$ are equal to zero. Because of our choice of $I_C$, no element of $I_C$ will satisfy any of the corresponding $k$ conditions, e.g. ``$\alpha_{i_1}=0$'',\ldots,``$\alpha_{i_k}=0$''. This amounts to consider $N-k$ instead of $N$ as the maximum number of conditions that one element of $I_C$ can satisfy in the previous reasoning, so that: 
\[
\hbox{card}\Big(L_{\vett{\alpha}\vett{\beta}}^a\Big)=
\begin{cases}
\binom{N-k}{a}g_a=\binom{N-k}{a}2^{N-a} & 0\leq a\leq N-k\\
0 & N-k<a\leq N
\end{cases};
\]
therefore in the present case $J_{\va\vb}=\frac{1}{3^{N-k}}\geq\frac{1}{3^{N}}$.
\hfill$\blacksquare$
\smallskip

\begin{note}
In the geometric picture, the set $I_C$ corresponds to the sub-hypercube of the lattice $L$ whose points have all coordinates greater than 0.
\end{note}

We are now able to construct bound entangled states in ${\cal S}_{2^N\times 2^N}$.
\begin{theorem}
The state $\rho_{I_{\rm BE}}(\vb_0)$, with $I_{\rm BE}=I_C\cup\{(\vett{0}_m,\vb_0)\}$, $\vb_0\neq0_n$ and $I_C$ given by \eqref{eq:pointdistribution}, is a PPTES.
\end{theorem}

\noindent
{\bf Proof}\quad The state is PPT because the sufficient and necessary conditon for positivity under PT of Theorem \ref{TH4} is satisfied. For this state $N_{I_{\rm BE}(\vb_0)}=3^N+1$. We use the result of Lemma \ref{TH5} with the slight difference that now all the points in $I_{\rm BE}(\vb_0)$ have a weight $\frac{1}{N_{I_{\rm BE}(\vb_0)}}=\frac{1}{3^N+1}$. Therefore for any $(\va,\vb)\in L$ the elements in $I_C$ contribute at least with $\frac{1}{N_{I_{\rm BE}(\vb_0)}}$ to $J_{\va\vb}$. On the other hand the element $(\vett{0}_m,\vb_0)\in I_{\rm BE}(\vb_0)$ contributes with $\pm\frac{1}{N_{I_{\rm BE}(\vb_0)}}$, the sign depending on the number of identical indices between $(\vett{0}_m,\vb_0)$ and $(\va,\vb)$. Therefore
\[
J_{\va\vb}\geq\frac{1}{N_{I_{\rm BE}(\vb_0)}}\pm\frac{1}{N_{I_{\rm BE}(\vb_0)}}\geq0.
\]
We check now that $\rho_{I_{\rm BE}(\vb_0)}$ is also entangled. In fact $\Sigma$ is a diagonal basis for the associated map $\Lambda_{\rho_{\pi}}$ of any LS $\rho_{\pi}$, as well as for the map $\Lambda_{\vb_0}$ of \eqref{eq:indecmap}. The eigenvalues of $C_{\Lambda_{\rho_{\pi}}}$ are $2^N\pi_{\va,\vb}$, $(\va,\vb)\in L$; in particular for  $\rho_{I_{\rm BE}(\vb_0)}$ they are
\[
\begin{cases}
\frac{2^N}{3^N+1} & (\va,\vb)\in I_{\rm BE}(\vb_0) \\
0			     & (\va,\vb)\notin I_{\rm BE}(\vb_0)
\end{cases},
\]
while those of $\Lambda_{\vb_0}$ are listed in \eqref{eq:evLambda}. Therefore in the case of $\rho_{I_{\rm BE}(\vb_0)}$, we have
\[
D_{\Lambda}(\rho_{I_{\rm BE}(\vb_0)})=\frac{1}{(2^N)^2}\sum_{(\va,\vb)\in I_{\rm BE}(\vb_0)}\frac{2^N}{3^N+1}\lambda_{\va\vb}=
	-\frac{1}{2^N(3^N+1)}<0.
\]
According to Proposition \ref{th:proposindec}, $\rho_{I_{\rm BE}(\vb_0)}$ is entangled and $\Lambda_{\vb_0}$ is not decomposable.\hfill$\blacksquare$
\medskip

\begin{figure}
\label{fig:PPTELS}
\centering
\subfigure[$m=1,\,n=1,\,N=2$]{\includegraphics[scale=0.4]{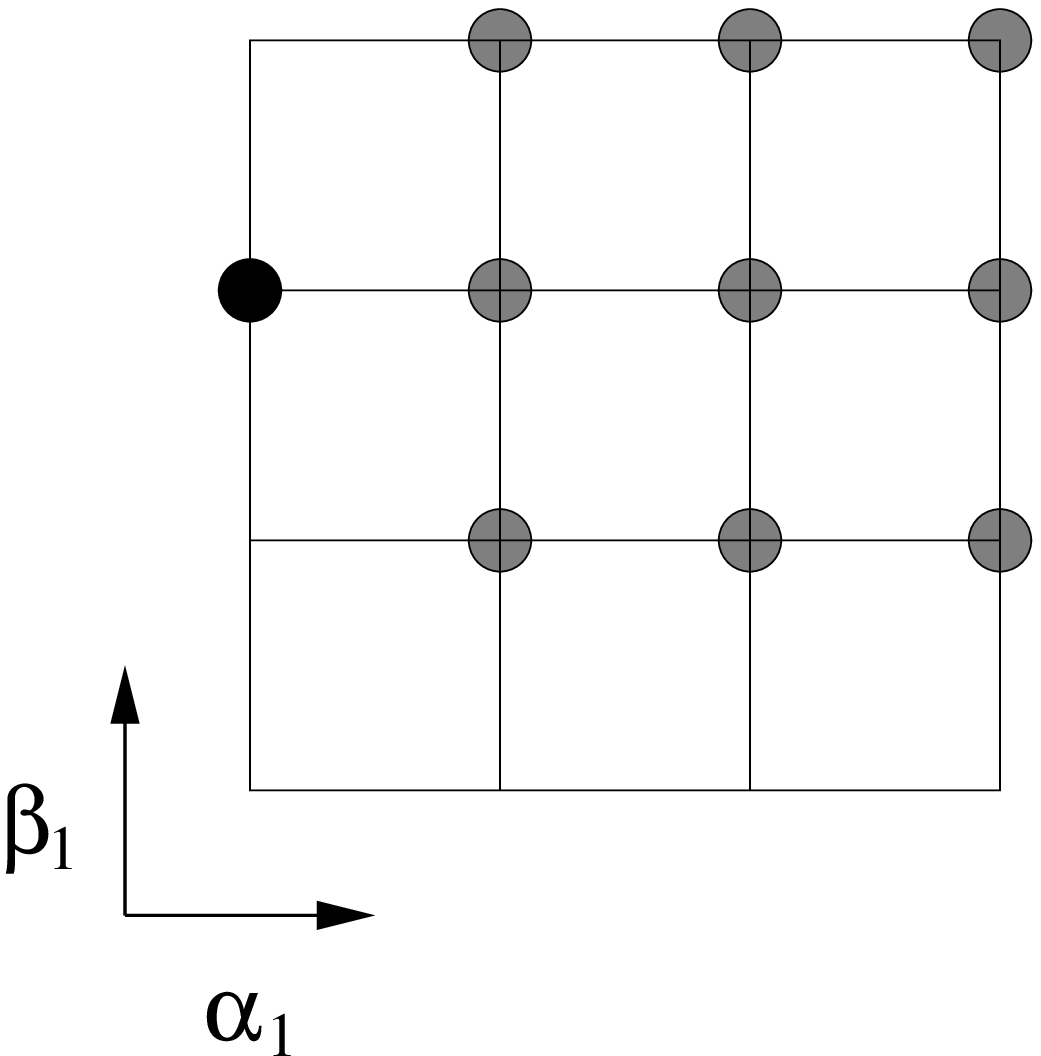}}
\hspace{0.7cm}
\subfigure[$m=2,\,n=1,\,N=3$]{\includegraphics[scale=0.4]{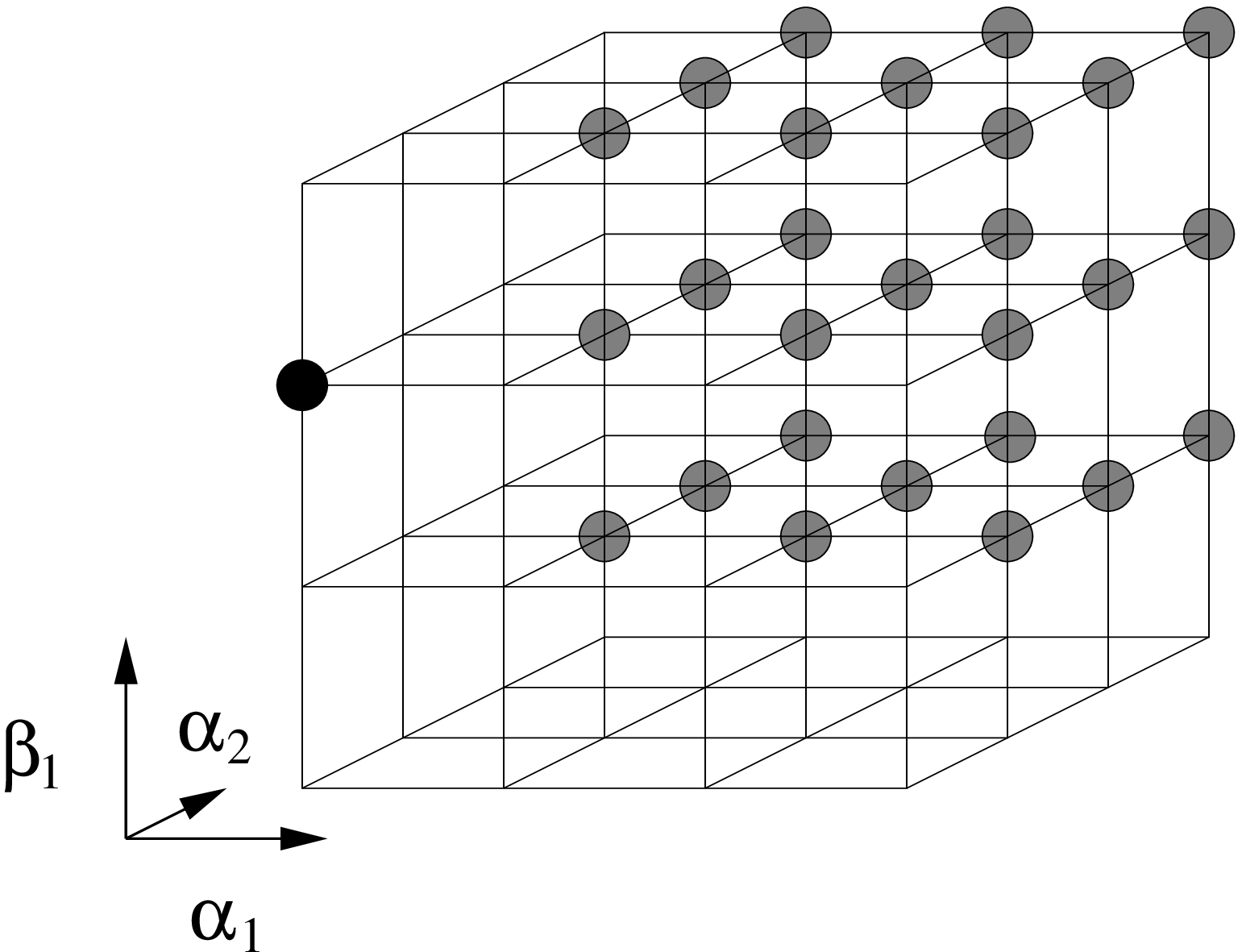}}
\caption{Geometric representation of two sets $I_{\rm BE}(\vb_0)$ identifying PPT entangled equidistributed lattice states: (a) $m=1$, $n=1$, $N=2$, $\vb_0=(2)$; (b) $m=2$, $n=1$, $N=3$, $\vb_0=(2)$. In both cases the black dot corresponds to the element  $(\vett{0}_m,\vb_0)$.}
\end{figure}

Two examples of sets $I$ describing PPT entangled ELS for N=2 and N=3 are shown in Figure \ref{fig:PPTELS}.

\begin{note}
Notice that local unitary operations preserve the properties of any state as regards entanglement and positivity under PT. It is therefore quite evident that our ``construction'', i.e. our choice for $I_C$, is just one of the possible. Generalizing the reasoning in \cite{BFPosid}, we note that, given two $2^N\times2^N$ unitary matrices $U,\,W$ such that $W\sigma_{\va\vb}U^\dagger=\sigma_{\vg\vd}$ up to a phase, we have
\begin{multline}
\big(U^*\otimes W\big)\,P_{\va\vb}\,\big(U^T\otimes W^\dagger\big)\\
=\Big[\idmat_{2^N}\otimes \big(W\sigma_{\alpha\beta}U^\dagger\big)\Big]
P^{2^N}_+
\Big[\idmat_{2^N}\otimes \big(U\sigma_{\alpha\beta}W^\dagger\big)\Big]
=P_{\gamma\delta}.
\end{multline}
Since the transformation is unitary and thus invertible, it induces a permutation among the elements of $L$, so that $\rho_{\pi'}=\big(U^*\otimes W\big)\rho_\pi\big(U^T\otimes W^\dagger\big)$ is another LS with permuted eigenvalues. Let us indicate with $\eta_i,\,i=1,\ldots,N$ the elements of the vector $(\va,\vb)$. For example $U$ and $W$ can be chosen such that
\[
(\eta_1,\ldots,\eta_i,\ldots,\eta_N)\mapsto(\eta_1,\ldots,p(\eta_i),\ldots,\eta_N),
\]
with $p:\{0,1,2,3\}\rightarrow\{0,1,2,3\}$ a permutation, or such that
\[
(\eta_1,\ldots,\eta_i,\ldots,\eta_j,\ldots,\eta_N)\mapsto(\eta_1,\ldots,\eta_j,\ldots,\eta_i,\ldots,\eta_N).
\]
In the geometric picture the first operation amounts to exchanging two parallel $(N-1)$-dimensional hyperplanes, while the second one corresponds to exchanging two coordinates. 
Therefore every set
\[
J\big((\vg,\vd),(\vett{\mu},\vett{\nu})\big)=I(\vg,\vd)\cup\big\{(\vett{\mu},\vett{\nu})\big\}
\]
with
\[
I(\vg,\vd)=\big\{(\va,\vb)\big|\alpha_i\neq\gamma_i,\beta_j\neq\delta_j,i=1,\ldots,m,j=1,\ldots,n\big\}
\]
$(\vett{\mu},\vett{\nu})\notin I(\vg,\vd)\cup\big\{(\vg,\vd)\big\} $ corresponds to a PPTES since it can be transformed into a $\rho_{I_{\rm BE}(\vb_0)}$ by means of the ``elementary'' operations just described.
\end{note}

\section{Conclusions}

A general class of positive but not completely positive maps has been found. The decomposability of a representative map of such class has been studied exploiting the characterization of entanglement by means of linear maps:  we have at the same time established the non-decomposability of the map and found examples of $2^N\times2^N$ dimensional states, e.g. states of a bipartite N+N qubits system, which are PPT but nevertheless entangled. Such examples are indeed interesting to analyse the phenomenon of bound entanglement.

The author thanks F. Benatti and R. Floreanini for fruitful discussions.


\end{document}